# A Multi Level Data Fusion Approach for Speaker Identification on Telephone Speech

Imen Trabelsi and Dorra Ben Ayed

*Electrical Engineering Department, National Engineering School of Tunis
Signal, Image and Pattern Recognition Research Unit,
Université Tunis El Manar BP 37, le Belvédère, Tunis 1002, Tunisia
trabelsi.imen1@gmail.com, Dorra.mezghani@isi.rnu.tn*

## Abstract

*Several speaker identification systems are giving good performance with clean speech but are affected by the degradations introduced by noisy audio conditions. To deal with this problem, we investigate the use of complementary information at different levels for computing a combined match score for the unknown speaker. In this work, we observe the effect of two supervised machine learning approaches including support vectors machines (SVM) and naïve bayes (NB).*

*We define two feature vector sets based on mel frequency cepstral coefficients (MFCC) and relative spectral perceptual linear predictive coefficients (RASTA-PLP). Each feature is modeled using the Gaussian Mixture Model (GMM). Several ways of combining these information sources give significant improvements in a text-independent speaker identification task using a very large telephone degraded NTIMIT database.*

## 1. Introduction

The task of automatic speaker identification consists of labeling an unknown voice sample as one of a set of known voices samples. Speaker identification methods can be divided into text-dependent and text-independent methods. For text dependent methods, the test utterance is known while for text independent methods, it is not known [9]. In this study, we performed experiments for text independent tasks in phone quality speech.

In general, speaker identification system is considered as comprised two major units namely enrollment (modeling) and identification (matching) as shown in (Figure 1). In enrollment phase, all samples from speakers are trained and stored in a database. The goal of enrollment is to construct a model for each speaker based on the features extracted from his/her speech samples. The identification is a process of computing a matching score between the input speech feature vector and a model of the speaker's voice. Success in speaker identification tasks depends on these two phases. Nowadays, and in order to improve the identification performance, attention has turned to the applicability of fusing different systems at different levels.

This technique can be divided in two main categories: systems based on features diversity [18] and systems based on classifiers diversity [19]. Hence, searchers are looking for the best set of features and the best set of classifiers.

In this paper, we study these two categories. It consists on comparing the use of individual classifiers and combined classifiers, applied in individual data and in combined data. This task is of a high interest in vocal applications through the telephone network. We use for our experiments the NTIMIT database [12]. This database represents a challenge due to the unpredictable channel noises.

The remainder of the paper is organized as follows. Section 2 reviews the basic of the gaussian mixture speaker model. Section 3 defines the structure of the used classifiers. Section 4 describes the application of the data fusion for text independent speaker identification. Section 5 describes the experimental conditions and presents the experimental results. Conclusion is drawn in the final section.

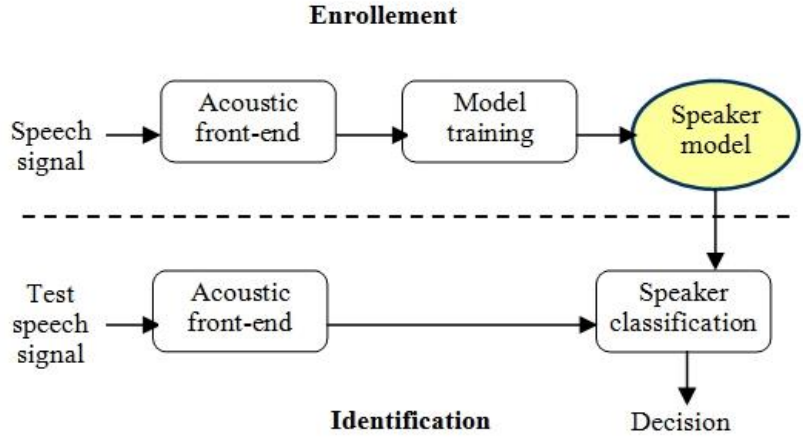

**Figure 1. General diagram of speaker identification system**

## 2. Speaker Modeling

The feature vectors extracted from training data are used to create a set of speaker models. The modeling of a speaker may be implemented according to various techniques. The widely used one is the Gaussian Mixture Model (GMM).

A GMM aims to approximate a complex nonlinear distribution using a mixture of simple gaussian models. A GMM is a weighted sum of M component Gaussian densities as given by (1):

$$p(x / \lambda) = \sum_{i=1}^{M} p_i \, b_i(x). \tag{1}$$

Where x is a d-dimensional vector, bi (x) are the component densities, pi are the mixture weights. Each component density is a d-variant Gaussian function having the form:

$$b_i(x) = \frac{1}{(2\Pi)^{\frac{d}{2}} \left| \sum_i \right|^{\frac{1}{2}}} \exp\{-\frac{1}{2}(x-\mu_i)^t \sum_i^{-1}(x-\mu_i). \tag{2}$$

With mean vector μi and covariance matrix $\sum$i, the mixture weights satisfy the constraint that:

$$\sum_{i=1}^{M} p_i = 1. \tag{3}$$

In this approach, a universal background model (UBM) learns the acoustic feature space. A standard approach in estimating the parameters of GMM-UBM aims to learn the mean vector, covariance matrix and the weight through expectation-maximization (EM) algorithm [13] from a background dataset.

Each speaker is then modeled and referred by adapting only the mean vectors of UBM using maximum a posteriori (MAP) criteria [5], while the weights and covariance matrix were set to the corresponding parameters of the UBM. All gaussian means vectors are pooled together to get one GMM supervector [3]. We produced GMM supervector on a per sequence. The GMM supervector can be thought of as a mapping between an utterance and a high-dimensional vector.

## 3. Classifiers

This section gives below a brief description of two supervised learning methods used to classify the GMM speaker models, namely NB and SVM.

### 3.1. Naïve Bayes (NB)

The naive bayes classification model is a simple classification technique based on Bayes' theorem. This classifier assumes that the effect of an attribute value on a given class label is independent of the values of the other attributes. This assumption is called class conditional independence. This classifier simply computes the conditional probabilities of the different classes given the values of attributes and then selects the class with the highest conditional probability.

If an instance is described with n attributes $a_i$ (i=1…n), then the class that instance is classified to a class c from set of possible classes C according to a Maximum a Posteriori (MAP) Naive Bayes classifier is:

$$c = \arg \max_{c_j \in C} \; p(c_j) \prod_{i=1}^{n} p(a_i \, (c_j)) \qquad (4)$$

Despite its simplicity, Naive Bayes can often outperform more sophisticated classification methods. For speaker recognition, NB algorithm is largely used, such in [8].

In our work, we propose the combination of both methods GMM and NB. We present the NB classifier in the supervector GMM space. This method is a novel approach substantially different from existing techniques.

### 3.2. Support Vector Machines (SVM)

Support vector machines are a new technique of the statistical learning theory proposed by Vapnick in 1995 [6] and are based on the structural risk minimization principle [11] from computational learning theory. SVMs are a new kernel method suitable for binary classification tasks that makes its decisions by constructing a hyperplane that optimally separates two classes.

In order to tackle non-linear classification problems, the input feature space is typically transformed to a higher dimensional space via a kernel function, where it is possible to define a hyperplane to separate both classes with maximum margin.

For speaker recognition, the first approach in using SVM classifier was implemented by Schmidt where SVM were trained directly on the acoustic space [6]. Another approach became recently more popular; consist of using an hybrid GMM-SVM [1, 2] in a way that the robustness advantage of generative models of GMM is combined with the discriminative power of SVM.

## 4. Multi Level Data Fusion

Data fusion techniques encompass any area which deals with applying a combination of different sources of information. The essential idea is to more accurately estimate the classification results via fusing the final outcomes of different experts (features, classifiers…) in an efficient way. Then the output of such combinations can be superior to all the individual experts.

### 4.1. Feature Level Fusion

**4.1.1. Vector Concatenation:** The idea of fusing different sources of features in speaker identification is not new. A well-known data fusion strategy is to concatenate the cepstral vectors with their delta (Δ) and delta-delta (ΔΔ) cepstra into a single feature vector, as shown in (Figure 2).

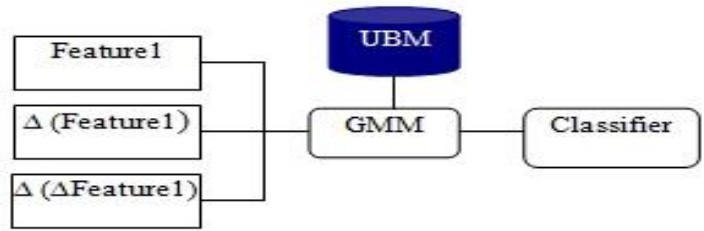

**Figure 2. Scheme of vector concatenation**

In general, vector concatenation is termed as classifier input fusion [10].

**4.1.2. GMM Supervectors Level Fusion:** We propose another multi input fusion by concatenating the GMM supervectors from features. In this case, we concatenate the outputs of modeling stage as summarized in (Figure 3).

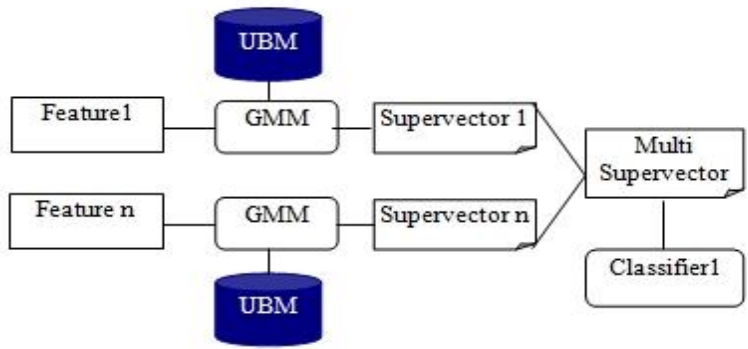

**Figure 3. Scheme of supervector fusion**

The multi-supervector is then presented to the speaker-independent classifier for scoring.

## 4.2. Scores Level Fusion

Researches in recent years show that fusion can be done in different levels and the score level fusion is the best in sense of simplicity and amount of information which supposed to be combined. The basic idea behind classifier combination is that when making a decision, one should not rely only on a single classifier, but rather classifiers need to participate in decision making by combining their individual scores.

In our case, each data is treated separately and each classifier independently makes its decisions as shown in (Figure 4).

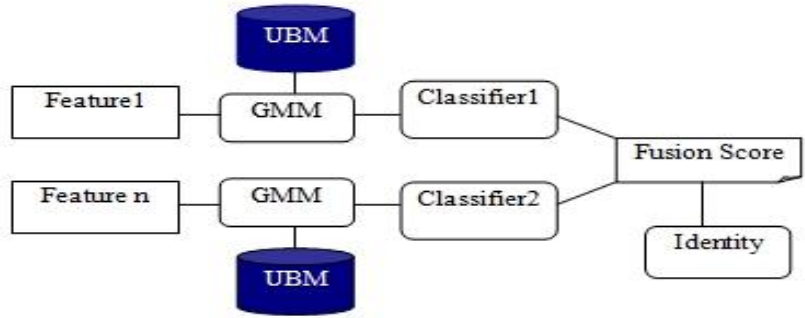

**Figure 4. Scheme of score fusion**

# 5. Experiments and Results

## 5.1. Corpus Description

The NTIMIT database is made of TIMIT utterances which are transmitted over a variety of telephone lines conditions. Each sentence of a speaker is transmitted over a different telephone line in order to have realistic conditions. All the utterances are different across speakers except utterance 'SA1' and 'SA2', which are common. We performed experiments on the dialect DR1 (New England dialect) for 28 speaker (14 female and 14 males). Each speaker spoke ten utterances, eight utterances are selected for training and two utterances are selected for the identification test. The data was digitized using 16 kHz sampling frequency with 16 bits per sample.

## 5.2. Feature Extraction

The speech feature extraction (front-end) is a key focus in robust speech recognition which significantly affects the recognition performance.

Each feature vector was extracted from pre-emphasized speech at 8ms intervals using a 16 ms window. A Hamming window was then applied to the speech frame.

In this study, we utilize mel-scale frequency cepstral coefficient (MFCC), these coefficients have been used as a standard acoustic feature set for identification system. Actually MFCC coefficients are the less vulnerable coefficient to noise perturbation [15].

We also use relative spectral perceptual linear predictive (RASTA-PLP) coefficients, this choice is emphasized by the fact that these coefficients are proved to be more robust for noisy environment [4] and have shown good recognition performance in a previous study [16].

We furthermore added the dynamic cepstrum parameters due to their popularity in automatic speaker recognition systems [17].

In our work, we construct 5 features data:

- Feature 1: characterized by the use of 12 MFCC coefficients.

- Feature 2: characterized by the use of 13 RASTA-PLP.

- Feature 3: characterized by the use of 12 MFCC coefficients accompanied by Delta and delta (Delta) coefficients, which sums to a 36 features.

- Feature 4: characterized by the use of 13 RASTA-PLP coefficients accompanied by Delta and Delta (Delta) coefficients, which sums to a 39 features.

- Feature 5: characterized by the concatenation of the supervectors issued from feature 1 and feature 2.

### 5.3. Baseline System

Our baseline system consists of a 128 component GMM-UBM built using the training utterances of all the 28 speakers. Individual speaker models are MAP-adapted; only mean vectors, with a relevance factor of 16.

After modeling speaker according to each feature data, we use individually the SVM and the NB for scoring and as a final step; we combine for each feature these two classifiers.

We construct three systems:

- System 1: SVM classifier.

- System 2: NB classifier.

- System 3: combination of SVM and NB classifiers.

To train SVM, we use a new learning algorithm called sequential Minimal Optimization (SMO) [14]. SVM algorithm was evaluated using linear kernel given by (5).

$$k(x, v_i) = x.v_i \ .$$
(5)

Where x is the input data and vi are the support vectors.

To handle with the SVM multi class, we are considered the one versus one approach. The SVM algorithm is implemented using the library LIBSVM [7].

Only for SVM, data were scaled to [0, 1] before the classification process.

The performance is measured as the identification rate (IR).

$$IR(\%) = \frac{Number \ of \ correct \ assignments}{Number \ of \ total \ assignments}.$$
(6)

### 5.4. Results

The results of the individual features are shown in (Table 1) for the three systems.

**Table 1. Results of individual and combined classifiers for individual features**

| | Identification rates IR (%) | | |
|---|---|---|---|
| **Feature Type** | **System1** | **System 2** | **System 3** |
| Feature 1 | 50 | 42.85 | 64,28 |
| Feature 2 | 53.57 | 50 | 75 |

Table 1 shows that the Identification rate is varied between (50%) and (75%).

We note that Feature 2 (RASTA-PLP) outperform Feature1 (MFCC) for the three systems. We also note that System1 (SVM) performs better than System2 (NB). The table shows also that the maximum IR is generated by combined classifiers for the two sets of feature.

The results of the combined features are shown in (Table 2) for the three systems.

**Table 2. Results of individual and combined classifiers for combined features**

| | Identification rates IR (%) | | |
|---|---|---|---|
| **Feature Type** | System1 | System 2 | System 3 |
| Feature 3 | 60,71 | 53,57 | 71,42 |
| Feature 4 | 57,14 | 50 | 71,42 |

Table 2 shows that the Identification rate is varied between (50%) and (71, 42%). For Feature 3, results prove that SVM classifier (System 1) performs better (60, 71%) than NB classifier (System 2) (53, 57%). For Feature 4, results prove that SVM classifier (System 1) performs better (57, 14%) than NB classifier (System 2) (50%). As observed, fusion score in System 3 improves accuracy, it provides the best IR equal to (71, 42%) for the two features data.

The performance of fusing supervectors for the three tested systems is evaluated in (Table 3).

**Table 3. Results of combining supervectors for individual and combined classifiers**

| | Identification rates IR (%) | | |
|---|---|---|---|
| **Feature Type** | **System1** | **System2** | **System3** |
| Feature 5 | 67,85 | 60,71 | 85,71 |

Table 3 shows that the Identification rate is varied between (60, 71%) and (85, 71%). It is noted that for the three systems, Feature 5 improve accuracy, the highest recognition accuracy is obtained using the System3 (85, 71%). The combination of the two feature type GMM supervectors shows a promising result.

## 6. Conclusion

In this paper, we have proposed several methods for text speaker identification which is very difficult through the telephone network, by fusing different feature extraction methods and different classifiers in order to improve the recognition accuracy.

The best recognition results are obtained from the concatenation of GMM supervectors from Rasta-PLP features and MFCC features with the use of combined classifiers.

In further work, we will try to study the performance of the proposed systems on the whole NTIMIT corpus.

## References

[1] S. Z. Boujelbene, D. B. A. Mezghani and N. Ellouze, "Robust Text Independent Speaker Identification Using Hybrid GMM-SVM System", Journal of Convergence Information Technology – JDCTA, vol. 3.2, ISSN: 1975-9339, **(2009)**, pp. 103-110.

[2] I. Trabelsi and D. B. Ayed, "Evaluation d'une approche hybride GMM-SVM pour l'identification de locuteurs", La revue e-STA, vol. 8, no. 1, **(2011)**, pp. 61-65.

[3] W. M. Campbell, D. E. Sturim, D. A. Reynolds and A. Solomonoff, "SVM based speaker verification using a GMM-supervector kernel and NAP variability compensation", Proc. Int. Conf. Acoustics, Speech, and Signal Processing, **(2006)**.

[4] H. Hermansky, N. Morgan, A. Bayya and P. Kohn, "Rasta plp speech analysis", International Computer Science Institute, 1947 Center Street; Berkeley, CA 94704,TR-91-069.

[5] D. Reynolds, T. Quatieri and R. Dunn, "Speakerverification using adapted gaussian mixture models", DSP, vol. 10, no. 3, (2000), pp. 19–41.

[6] M. Schmidt and H. Gish, "Speaker Identification via Support Vector Machies", in ICASSP, **(1996)**, pp. 105-108.

[7] C. -C. Chang and C. -J. Lin, "LIBSVM: a library for support vector machines", **(2001)**, http://www.csie.ntu.edu.tw/~cjlin/libsvm.

[8] L. Toth, A. Kocsor and J. Csirik, "On naive bayes in speech recognition", nt. J. Appl. Math. Comput. Sci., vol. 15, no. 2, **(2005)**, pp. 287–294.

[9] D. Reynolds, "An Overview of Automatic Speaker Recognition Technology", the International Conference on Acoustics, Speech, and Signal Processing ICASSP 02, Orlando, Florida, USA, **(2002)**, pp. 4072–4075.

[10] S. Slomka, S. Sridharan and V. Chandran, "A Comparison of fusion techniques in mel-cepstral based speaker idenficication", Proc. ICSLP 1998, Sydney, Australia, **(1998)**.

[11] V. Vapnik, "Statistical learning theory", S. Haykin, (Ed.), Adaptive and Learning Systems for Signal Processing, Communications, and Control, John Wiley & Sons, **(1998)**.

[12] W. Fisher, V. Zue, J. Bernstein and D. Pallet, "An Acoustic-Phonetic Data Base", J. Acoust. Soc. Amer. Suppl. (A), vol. 81, S92, **(1986)**.

[13] A. P. N. Dempster, M. D. Laid and B. Durbin, "Maximum Likelihood from incomplete data via the EMalgorithm", J. Royal Statistical Soc., vol. 39, **(1977)**, pp. 1-38.

[14] J. C. Platt, "Sequential minimal optimization: A fast algorithm for training support vector machines", Rapport interne, Microsoft Research, **(1998)**.

[15] S. B. Davis and P. Mermelstein, "Comparison of Parametric Representation for Monosyllabic Word Recognition in Continuously Spoken Sentences", IEEE Trans. On ASSP, vol. ASSP 28, no. 4, **(1980)** August, pp. 357-365.

[16] I. Trabelsi and D. Ben Ayed, "Stratégies de fusion de paramètres pour une tâche d'identification du locuteur en mode indépendant du texte en environnement bruité: Application sur le corpus Ntimit", Taima'11, Hammamet, Tunisie, **(2011)**.

[17] F. K. Soong and A. E. Rosenberg, "On the Use of Instantaneous and Transitional Spectral Information in Speaker Recognition", IEEE Trans. Acoustics, Speech and Signal Processing, vol. 36, no. 6, **(1988)**, pp. 871-879.

[18] E. Monte-Moreno, M. Chetouani, M. Faundez-Zanuy and J. Sole-Casals, "Maximum Likelihood Linear Programming Data Fusion for Speaker Recognition", Speech Communication, vol. 51, no. 9, **(2009)**, pp 820-830.

[19] J. Kittler, M. Hatef, R. P. W. Duin and J. Matas, "On Combining Classifiers", IEEE Trans. Pattern Analysis and Machine Intelligence, vol. 20, no. 3, **(1998)**, pp. 226-239.

# Authors

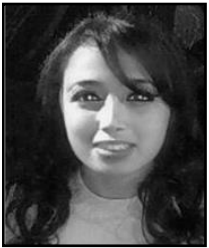

**Imen Trabelsi**

I. Trabelsi received a university diploma in computer science in 2009 from the High Institute of Management of Tunis (ISG-Tunisia), the MS degree in signal processing in 2011 from the *Institute of Computer Science of Tunis (ISI-Tunisia).* She is currently working towards the Ph.D. degree in electrical engineering (signal processing) at the *National School of Engineer of Tunis (ENIT).* Her areas of interests are speech processing, pattern recognition, emotion recognition and speaker recognition

E-mail: trabelsi.imen1@gmail.com

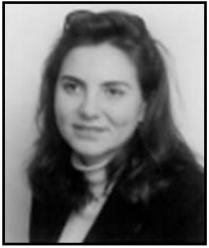

**Dorra Ben Ayed  Mezghani**

D. Ayed Mezghani received computer science engineering degree in 1995 from the National School Computer Science (ENSI-Tunisia), the MS degree in electrical engineering (signal processing) in 1997from the National School of Engineer of Tunis (ENITTunisia), the Ph. D. degree in electrical engineering (signal processing) in 2003 from (ENIT-Tunisia).

She is currently an associate professor in the computer science department at the High Institute of Computer Science of Tunis (ISI-Tunisia). Her research interests include fuzzy logic, support vector machines, artificial intelligence, pattern recognition, speech recognition and speaker identification.

E-mail: Dorra.mezghani@isi.rnu.tn, DorraInsat@yahoo.fr